\documentclass[twocolumn,twocolappendix]{aastex63}

\usepackage[utf8]{inputenc} 

\usepackage{graphicx} 
\usepackage{amssymb,amsmath}

\usepackage{array} 

\usepackage{units}
\usepackage{subfigure}
\usepackage{stackengine}

\shorttitle{Torus instability threshold}
\shortauthors{Alt et al.}

\graphicspath{{./}{figures/}}

\newcommand{\pD}[2]{\frac{\partial #2}{\partial #1}}

\newcommand{\D}[2]{\frac{{\rm d} #2}{{\rm d} #1}}

\newcommand\bb[1]{\mbox{\boldmath{$#1$}}}

\newcommand\btimes{\bb{\times}}

\newcommand{\mrm}[1]{\mathrm{#1}}
\newcommand{\ms}[1]{\mrm{#1}} 
\newcommand{\Eq}{equation~}
\newcommand{\Fig}{Figure~}

\newcommand{\ex}{\ensuremath{\hat{\bb{x}}}}

\newcommand{\const}{{\rm Const.}}

\newcommand{\Fh}{\ensuremath{F_\ms{h}}}
\newcommand{\Ft}{\ensuremath{F_\ms{t}}}
\newcommand{\Fs}{\ensuremath{F_\ms{s}}}

\newcommand{\FN}{\ensuremath{F_\ms{N}}}
\newcommand{\FNdef}{$\FN\equiv\mu_0\It^2/(4\pi\xf)$}
\newcommand{\ndef}[1][]{$n_\ms{#1}=-z\,\partial\,(\ln{B_\ms{#1}})/\partial z$}
\newcommand{\ns}{\ensuremath{n_\ms{s}}}
\newcommand{\ncr}[1][]{\ensuremath{n_{\ms{cr}}^\ms{#1}}}
\newcommand{\ncrLab}{\ncr[lab]}
\newcommand{\ncrSol}{\ncr[solar]}
\newcommand{\qcr}[1][]{\ensuremath{q_{\ms{cr}#1}}}
\newcommand{\qa}{\ensuremath{q_a}}
\newcommand{\qmax}{\ensuremath{q_\ms{max}}}
\newcommand{\xf}{\ensuremath{x_\ms{f}}}
\newcommand{\It}{\ensuremath{I_\ms{T}}}
\newcommand{\Bs}{\ensuremath{B_\ms{s}}}
\newcommand{\Bg}{\ensuremath{B_\ms{g}}}
\newcommand{\Bpi}{\ensuremath{B_\ms{Pi}}}
\newcommand{\Bti}{\ensuremath{B_\ms{Ti}}}

\newcommand{\BP}{\ensuremath{B_\ms{P}}}
\newcommand{\BPa}{\ensuremath{B_{\ms{P},a}}}
\newcommand{\JT}{\ensuremath{J_\ms{T}}}
\newcommand{\JP}{\ensuremath{J_\ms{P}}}
\newcommand{\zapex}{\ensuremath{z_\ms{ap}}}

\newcommand{\popRef}{\citet{myers2016}}
\newcommand{\citec}[1]{\citep{#1}} 
\newcommand{\PPPL}{Princeton Plasma Physics Laboratory, PO Box 451, Princeton, NJ 08543, USA}
\newcommand{\PU}{Department of Astrophysical Sciences, Princeton University, Peyton Hall, Princeton, NJ 08544, USA}
\newcommand{\Postdam}{Institute of Physics and Astronomy, University of Potsdam, Potsdam D-14476, Germany}
\newcommand{\CfA}{Harvard-Smithsonian Center for Astrophysics, 60 Garden Street, Cambridge, MA 02138, USA}
\newcommand{\Sandia}{Sandia National Laboratories, 1515 Eubank SE, Albuquerque, New Mexico 87185, USA}

\usepackage{verbatim}

\submitjournal{ApJ}

\begin{document}

\title{Laboratory study of the torus instability threshold in solar-relevant, line-tied magnetic flux ropes}

\correspondingauthor{Andrew Alt}
\email{aalt@pppl.gov}

\author[0000-0001-9475-8282]{Andrew Alt}
\affiliation{\PU}\affiliation{\PPPL}
\author[0000-0003-4539-8406]{Clayton E. Myers}
\affiliation{\Sandia}
\author[0000-0001-9600-9963]{Hantao Ji}
\affiliation{\PU}\affiliation{\PPPL}
\author[0000-0003-0760-6198]{Jonathan Jara-Almonte}
\affiliation{\PPPL}
\author[0000-0003-3881-1995]{Jongsoo Yoo}
\affiliation{\PPPL}
\author[0000-0001-8093-9322]{Sayak Bose}
\affiliation{\PPPL}
\author[0000-0003-3639-6572]{Aaron Goodman}
\affiliation{\PU}\affiliation{\PPPL}
\author[0000-0003-4996-1649]{Masaaki Yamada}
\affiliation{\PPPL}
\author[0000-0002-5740-8803]{Bernhard Kliem}
\affiliation{\Postdam}
\author[0000-0002-5598-046X]{Antonia Savcheva}
\affiliation{\CfA}


\begin{abstract}
    Coronal mass ejections (CME) occur when long-lived magnetic flux ropes (MFR) anchored to the solar surface destabilize and erupt away from the Sun. This destabilization is often described in terms of an ideal magnetohydrodynamic (MHD) instability called the torus instability. It occurs when the external magnetic field decreases sufficiently fast such that its decay index, \ndef{}, is larger than a critical value, $n>\ncr$, where $\ncr=1.5$ for a full, large aspect ratio torus. However, when this is applied to solar MFRs, a range of conflicting values for \ncr{} is found in the literature. To investigate this discrepancy, we have conducted laboratory experiments on arched, line-tied flux ropes and have applied a theoretical model of the torus instability. Our model describes an MFR as a partial torus with footpoints anchored in a conducting surface and numerically calculates various magnetic forces on it. This calculation yields a better prediction of \ncr{} which takes into account the specific parameters of the MFR. We describe a systematic methodology to properly translate laboratory results to their solar counterparts, provided that the MFRs have sufficiently small edge safety factor, or equivalently, large enough twist. After this translation, our model predicts that \ncr{} in solar conditions often falls near $\ncrSol\sim0.9$ and within a larger range of $\ncrSol\sim(0.7,1.2)$ depending on the parameters. The methodology of translating laboratory MFRs to their solar counterparts enables quantitative investigations of the initiation of CMEs through laboratory experiments. These experiments allow for new physics insights that are required for better predictions of space weather events but are difficult to obtain otherwise. 
\end{abstract}

\date{\today}


\section{Introduction}


Protrusions of magnetic field and plasma from the solar surface often result in the formation of long, thin magnetic flux ropes (MFR) \citec{kuperus1974,chen1989,rust2003}. These flux ropes are bundles of twisted magnetic field lines with footpoints anchored to the Solar surface through line-tying to the conductive photosphere. The ropes are often long-lived but can sometimes violently erupt, leading to coronal mass ejections (CME) \citec{crooker1997,green2009}. Understanding the causes of these eruptions is necessary for the prediction and further understanding of space weather.

One potential mechanism that can trigger a CME from an initially stable MFR is an ideal magnetohydrodynamic (MHD) instability called the torus instability \citec{bateman1978,kliem2006}. This instability occurs when the external magnetic field perpendicular to the MFR's axis, the strapping field, decays quickly enough with height. The rate of decay of the strapping field can be described by a decay index, \ns, and causes an instability when it exceeds a critical value, $\ns>\ncr$. 
For an axisymmetric, large aspect ratio, full torus the critical value is $\ncr=1.5$ \citec{shafranov1966,bateman1978}. However, extending this theory to non-toroidally symmetric, line-tied systems such as MFRs has proved challenging \citec{isenberg2007,olmedo2010}. While some MFR simulations are consistent with $\ncr=1.5$ \citec{torok2007,aulanier2009,zuccarello2015}, other analytical work has found critical values in a much larger range, $0.5<\ncr<2$ depending on the ratio of the apex height and footpoint half-separation \citec{olmedo2010}. Some simulations have found values near the higher end of this range with $1.75<\ncr<2$ \citec{fan2007,fan2010}. This simulated range is inconsistent with some observational evidence of $\ncr\sim1.3$ \citec{duan2019}. Critical values on the lower end of the range presented in \citet{olmedo2010} have been seen in recent laboratory experiments where an empirical threshold of $\ncr\sim0.8$ was seen \citec{myers2015}. This experimental value is also consistent with some solar observations \citec{jing2018}. 

One cause of the large discrepancy in previous values of \ncr{} comes from inconsistencies in definitions. For example, while some works define \ns{} at the rope's apex height, \zapex{} \citec{jing2018}, others choose a value above the rope even up to $z=2\zapex$ \citec{chen1996}. Others still define \zapex{} as the location where $n_s=\ncr$ for an assumed value of \ncr{} \citec{wang2017}. Due to difficulty of solar measurements, sometimes the decay index of the total external magnetic field is used rather than that of just the strapping field, which is used in the theory \citec{liu2008}. Practical measurement issues also cause some to measure \ns{} at a fixed height rather than at a dynamic one for each rope \citec{liu2008}. While the value of \ns{} at a fixed height is likely correlated with the true $\ns(\zapex)$, it cannot be used to determine \ncr{} or even be directly compared against it. With these considerations, it is no surprise that the literature contains many conflicting values and ranges for \ncr.

In order to reconcile the different values of \ncr{}, we devise a simple MFR model and use it to numerically calculate \ncr{} as a function of the flux rope parameters. The relevant dimensionless parameters are the ratio of the apex height to the footpoint half-separation, the aspect ratio of the partial torus, and the normalized internal inductance. The prediction can be applied to either laboratory or solar conditions to yield different results, \ncrLab{} and \ncrSol. These \ncr{} predictions are then validated against previous \citec{myers2015} and more recent experimental results. Our numerical results predict critical values in the range of $\ncrLab\sim(0.65,1.1)$ when applied to our experimental conditions. This is consistent with the previous empirical value of $\ncrLab\sim0.8$ but greatly increases our understanding of it and provides detailed dependence on experimental conditions and MFR parameters. Our model is able to predict erupting ropes in our experiments with a true positive rate of 94\%. This high detection rate indicates that our method could ultimately be used to improve space weather predictions.

We also present how these values should be translated to the conditions on the Sun, which makes the majority of our results clustered near $\ncrSol\sim0.9$ with a full range extending to $\ncrSol\sim(0.7,1.2)$. The results reported here can better explain the wide range of \ncrSol{} values reported in the literature. This new methodology of translating laboratory MFRs to their solar counterparts paves a systematic way to quantitatively investigate the initiation of CMEs through laboratory experiments. These experiments allow for new physics insights that are required for better predictions of space weather events but difficult to obtain otherwise.

In the rest of this paper, we briefly introduce the torus instability and the kink instability, another important MHD instability in arched, line-tied flux ropes. This is followed by the development of numerical models of each of the contributing forces and resulting critical decay index applicable to our laboratory experiments. The model allows for a more accurate prediction of \ncr{} than can normally be achieved though purely analytical means. The predicted values of \ncr{} also lie in a range that depends on the exact parameters of the MFR. The range allows for more accurate predictions than can be achieved by a single scalar value. The results of the experiments are then translated to the conditions on the Sun to create a better prediction of MFR eruptions and subsequent CMEs.

\section{Flux ropes and associated instabilities}

\begin{figure*}%
    \centering
    \subfigure{\label{fig:chen-model}}%
    \subfigure{\label{fig:MRX}}%
    \stackinset{l}{.5cm}{t}{.5cm}{\textbf{(a)}}{\subfigure{\includegraphics[width=.4\linewidth]{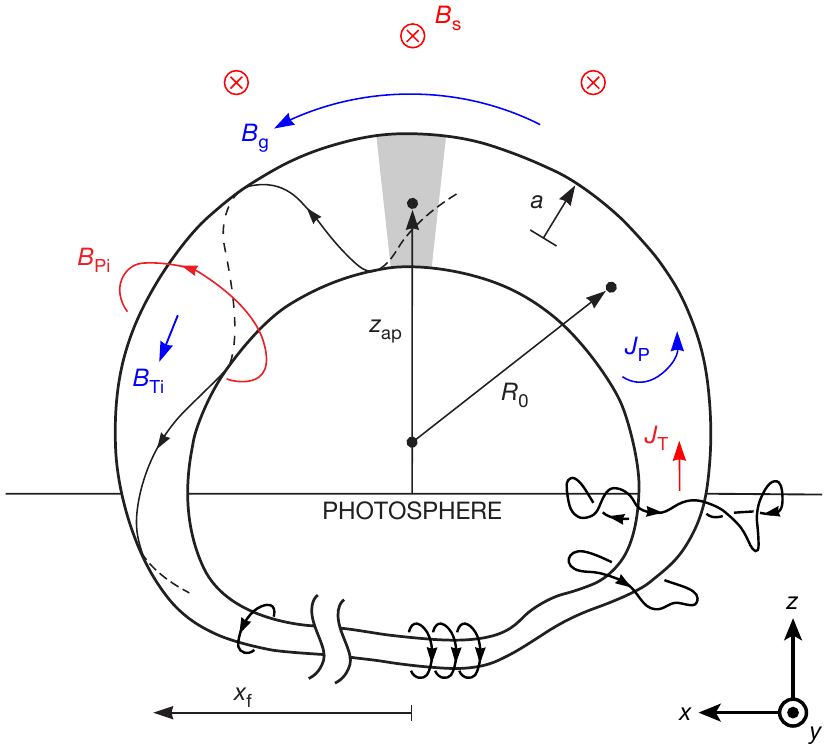}\label{fig:chen-model2}}\hfill}%
    \stackinset{l}{.5cm}{t}{.5cm}{\textbf{(b)}}{\subfigure{\includegraphics[width=.59\linewidth]{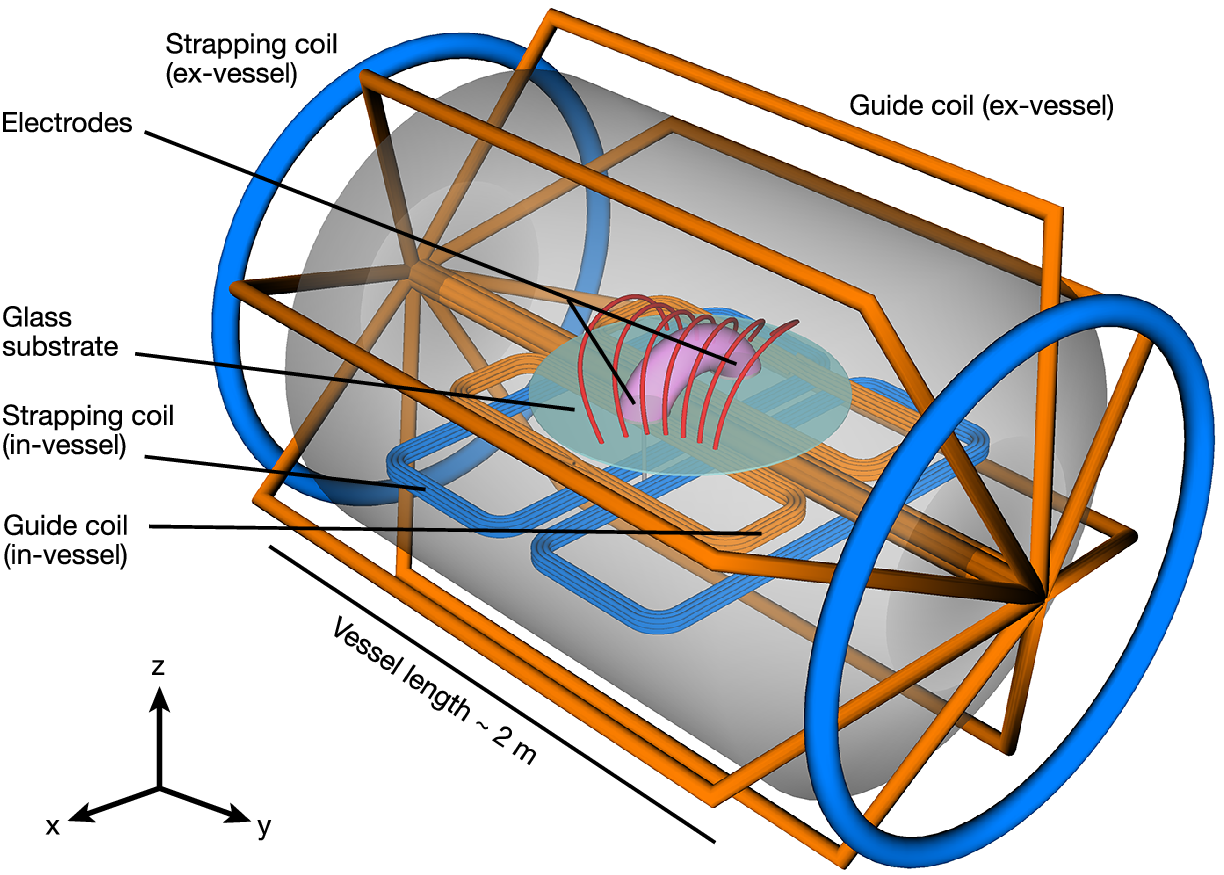}\label{fig:MRX2}
    }}
	\caption{\textbf{(a)} Breakdown of fields and currents in an arched, line-tied magnetic flux rope. The footpoints are anchored onto the conducting photosphere with a separation of $2\xf$. The external fields, \Bs{} and \Bg, are generated by the Sun while the internal fields, \Bpi{} and \Bti, are generated by the currents in the rope. Figure reproduced from \popRef{} and adapted from \citet{chen1989,chen2003}. %
	\textbf{(b)} The MRX vessel used to create arched, line-tied flux ropes. An arc discharge is created between two copper electrodes and is separated from the magnetic-field coils by a glass substrate. The model in (a) corresponds to the pink plasma arc in the center of the image. Four coils were inserted into MRX in order to control the profiles of both the guide and strapping fields. The orange coils contribute to the guide field along the rope while the blue coils control the strapping field across it. Control of the vacuum fields allows for control of the instability parameters for the torus and kink instabilities. Figure reproduced from \citet{myers2015}.}%
\end{figure*}

The magnetic field, \bb{B}, and electric currents, \bb{J}, of an MFR can be broken up into components based on their source and direction as shown in \Fig\ref{fig:chen-model}. First \bb{B} can be separated into external and internal components based on whether they are generated by currents in the Sun or currents within the MFR itself. The external field is further divided into the guide field, \Bg, along the axis of the MFR, and the strapping field, \Bs, perpendicular to this axis. The internal fields and the currents are then separated into toroidal, T, and poloidal, P, components. In a low-$\beta$ plasma (where $\beta\equiv2\mu_0 P/B^2$ is the ratio of thermal to magnetic energy in the plasma), the dominant force on an MFR is the $\bb{J}\btimes\bb{B}$ force. From the decomposition of \bb{J} and \bb{B}, this force can also be decomposed based on the source terms. These forces are called the hoop, strapping, and tension forces and are defined in Table~\ref{tab:forces}. The hoop force caused by the toroidal current interacting with the internal poloidal field causes the MFR to expand upward while the tension and strapping forces resist this expansion and hold it down.

\begin{deluxetable*}{lCCC}
	\caption{Breakdown of the magnetic forces on a flux rope\label{tab:forces}}
	\tablehead{
	\colhead{Force}  &  \colhead{Symbol}     & \colhead{Source Term} & \colhead{Analytical Expression} }%
 	\startdata
 	Hoop force (upward)   & \Fh      & $f_\ms{h}=\JT\Bpi$             & $\Fh=\frac{\mu_0\It^2}{4\pi R}\left[\ln\left(\frac{8R}{a}\right)-1+\frac{\ell_\ms{i}}{2}\right]$\tablenotemark{\footnotesize\rm{a}}%
 	\\ 
 	Strapping force (downward)   &  \Fs & $f_\ms{s}=-\JT\Bs$             & $\Fs=-\It\Bs$ \\ 
 	Tension force (downward)   &  \Ft   & $f_\ms{t}=-\JP(\Bg+\Bti)$ & $\Ft=\frac{-\mu_0\It^2}{8\pi R}\left[\frac{\left< B_\ms{T}^2\right>-B_\ms{g0}^2}{B_{\ms{P}a}^2} \right] \approx -\frac{1}{2}\frac{\mu_0\It^2}{4\pi R} $\tablenotemark{\footnotesize\rm b}\\ 
    \enddata
\tablecomments{The forces are separated based on their source terms in the $\bb{J}\btimes\bb{B}$ force. The fields and currents used are shown in \Fig\ref{fig:chen-model} and the analytical expressions are derived in detail by \popRef.}
\tablenotetext{\footnotesize\rm a}{ This expression for \Fh{} is different than some other sources due to the separation of \Ft.}
\tablenotetext{\footnotesize\rm b}{ This approximation is derived from minor radius force balance.}
\end{deluxetable*}

\subsection{Torus Instability Parameter}

The torus instability occurs when the net force on an MFR increases with height away from equilibrium, i.e.
\begin{equation}\label{eqn:TI_force_balance}
    \left.\sum_i F_i=0\right|_{z=\zapex} \quad \text{and} \quad \left.\sum_i\pD{z}{F_i}>0\right|_{z=\zapex} ,
\end{equation}
where $F_i$ are the constituent forces and \zapex{} is the equilibrium height of the rope's apex. The instability can therefore also be thought of as a loss of equilibrium where small perturbations cause the $\bb{J}\btimes\bb{B}$ forces to push the rope away from equilibrium. Since the main downward force is caused by the external strapping field, the torus instability criterion is often cast in terms of the decay index\footnote{The decay index is defined such that a field, $B\propto z^{-n}$ has a decay index of $n$.} %
of this field,
\begin{equation}
\ns=-\frac{z}{\Bs}\pD{z}{\Bs} > \ncr,
\end{equation}
where $z$ is the height above the photosphere and $\ncr=1.5$ in an axisymmetric, large aspect ratio, full torus. That is, the torus instability occurs when the strapping field decays too quickly with height above the photosphere.

\subsection{Kink Instability Parameter}
Another ideal MHD instability that can affect MFRs is the kink instability \citec{kruskal1954, shafranov1956, torok2004}. This instability occurs when the toroidal current, \It, in a flux rope is too large, or equivalently, the guide field, \Bg, is too small. When the current causes the outer field lines to fully twist around the rope, the instability can onset. The instability criterion can be described in terms of the edge safety factor, \qa, or equivalently, twist number, $T_\ms{w}$ 
\begin{equation}
\qa\equiv \frac{1}{|T_\ms{w}|} = \frac{2\pi a}{L}\frac{B_{\ms{T},a}}{\BPa}  < \qcr  ,
\end{equation}
where $a$ and $L$ are the minor radius and length of the MFR, $B_{\ms{T},a}$ is the edge toroidal magnetic field, $\BPa\equiv\mu_0\It/(2\pi a)$ is the edge poloidal field, and \qcr{} is the critical safety factor below which the instability occurs. In a toroidally symmetric system, such as a tokamak, the critical value is $\qcr=1$. However, in line-tied systems where there is no toroidal symmetry \qcr{} is modified \citec{ryutov2006}. In recent experiments with arched, line-tied MFRs, the critical value was found to be near $\qcr\sim0.8$ \citec{myers2016}. This value is also consistent with some analytical work \citec{hood1981}. 

\subsection{Torus and Kink Instabilities' Effect on MFR Eruptions}
\label{sec:TI-KI-effect}

Both the torus and kink instabilities are often considered when studying the onset of CMEs. However, since the kink instability quickly saturates, it is unlikely to be the sole cause of CMEs, and the torus instability is required for full eruptions \citec{torok2005}. The kink instability of MFRs has been extensively verified in laboratory experiments and will not be the focus of this paper \citec{hsu2003,bergerson2006,oz2011,ha2016}. 

While the torus instability is often described only in terms of \ns, recent experiments have found that ropes with both $\ns>\ncr$ and $\qa>\qcr$ were relatively stable \citec{myers2015,myers2017PPCF}. This ``failed torus" regime involves reorganization that stabilizes an otherwise erupting rope. Flux ropes that were torus unstable but ultimately confined have also been found in solar observations \citec{zhou2019}. Since ropes in this regime do not undergo the standard torus instability, they will not be the focus of this paper and we will limit most of our analysis to ropes with $\qa<\qmax$, where a value of $\qmax=0.67$ will be used and justified by arguments in Section~\ref{sec:statistics}.

\section{Quasi-analytical Flux Rope Model}\label{sec:model}

\begin{figure}
    \centering
	\includegraphics[width=\linewidth]{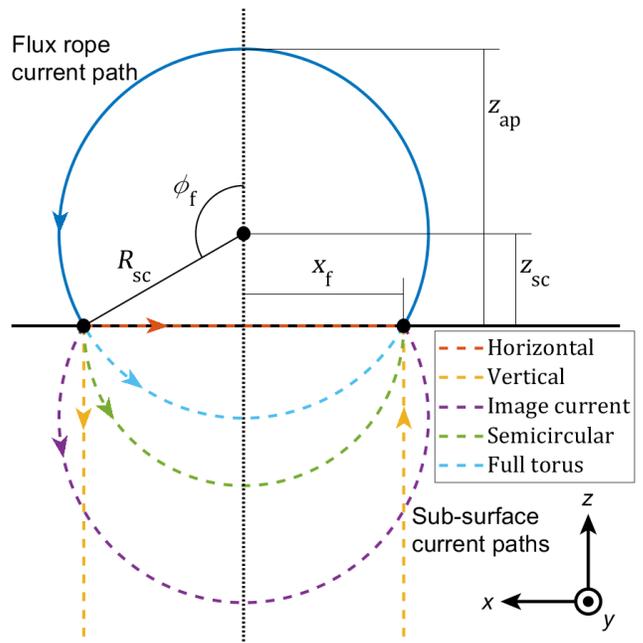}
	\caption{The shifted-circle model and different current closure options that are considered. The axis of the flux rope is approximated by the solid blue circle, which has footpoints fixed at $x=\pm\xf$ and apex height, \zapex. Given \xf{} and \zapex{}, the radius of curvature, $R_\ms{sc}$, center of curvature, $z_\ms{sc}$, and angle, $\phi_\ms{f}$, can be determined, making this a 1-D model with a single parameter, \zapex. The dashed curves represent different potential paths that the return current can take below the solar surface at $z=0$. The return currents (dashed lines) shown are either fixed paths or a dynamic image current (purple). The fixed paths are: horizontal closure between the footpoints (red), vertical paths off to $z\to-\infty$ (yellow), a fixed, semicircular path (green), and a full circle (cyan).}
	\label{fig:closureModels}
\end{figure}

A simple model of a solar flux rope is that of a partial torus with ends anchored in the solar surface (see \Fig\ref{fig:closureModels}). In order to approximate the path of a flux rope, we will use a ``shifted-circle" model such as in \citet{chen1989} and \popRef. In this model, the axis of a flux rope is assumed to follow the path of a circle with footpoints fixed on the solar surface, which is modeled as a perfectly conducting region for $z<0$. By keeping the location of the footpoints fixed, line-tying at the solar surface can be maintained while the flux rope height is changed. 

In this model, the radius of curvature for the MFR, $R_\ms{sc}$, and the height of its center of curvature, $z_\ms{sc}$, are given by
\begin{equation}
    R_\ms{sc} = \frac{\zapex^2+\xf^2}{2\zapex} \quad\text{and}\quad z_\ms{sc} = \frac{\zapex^2-\xf^2}{2\zapex},
\end{equation}
where \zapex{} is the apex height and \xf{} is the (fixed) footpoint half-separation. The angle from the $z$-axis to the footpoint, $\phi_\ms{f}$, which measures the fraction of the circle present above $z=0$, is given by
\begin{equation}
    \phi_\ms{f}  = 2\tan^{-1}\left(\frac{\zapex}{\xf}\right). 
\end{equation}
From these equations, we see that the shifted-circle model is one-dimensional, defined by one parameter, \zapex{}, and one constant, \xf{}.

However, the shifted-circle only describes the rope above the solar surface at $z=0$. In order for the model to be self-consistent, the toroidal current flowing through the rope must close. Different options for this closure are shown in \Fig\ref{fig:closureModels}. In the case of solar flux ropes, the most reasonable current closure is that of an image current in order to maintain the value of the normal magnetic field at the surface. However, in laboratory experiments the return current flows through wires and thus should be fixed and independent of \zapex.

\label{sec:ncr-parameters}
Another crucial part of modeling MFRs is a description of their rise and expansion. While the magnetic axis is defined by just \zapex{} and \xf, the hoop force depends on other parameters of the rope. These include the total toroidal current, \It, the inverse aspect ratio, $\varepsilon\equiv a/R$, and the internal inductance, $\ell_i\equiv\langle \BP^2\rangle/\BPa^2$, where $\langle B_\ms{P}^2\rangle$ is the cross-section average of $\BP^2$. The internal inductance measures the distribution of toroidal current. Specifically, $\ell_i=0.5$ for a uniform current and increases for current distributions that are more peaked around the axis. The evolution with respect to \zapex{} of these parameters during an instability must also be specified. The short-term evolution of $\It(\zapex)$ is often defined by the conservation of magnetic flux below the rope \citec{kliem2006,olmedo2010}. However, in the laboratory experiments described in Section~\ref{sec:exp-setup}, \It{} is fixed on the timescale of an eruption due to the large external inductance in the capacitor banks driving the ropes and so $\It(\zapex)=\const$ We will also consider a self-similar expansion of the ropes. This means, in part, that $\varepsilon(\zapex)=\const$ Since the hoop force is only logarithmically dependant on $\varepsilon$, it is not very sensitive to the choice of $\varepsilon(\zapex)$ and a self-similar expansion should be a reasonable approximation. The value of $\ell_i$ is determined by the radial distribution of toroidal current. Therefore, a self-similar expansion also implies that $\ell_i(\zapex)=\const$ However, $\ell_i$ does not fully describe the current distribution and a specific distribution must be chosen in order for numerical calculations to be carried out. As a model current distribution, we have used
\begin{equation}
    \JT(r)=\frac{\It}{\pi a^2}(1+\alpha) \left[1-\left(\frac{r}{a}\right)^2\right]^\alpha,
\end{equation}
where $r$ is the minor radius coordinate and $\alpha>-1$ is a free parameter that has a one-to-one correspondence to $\ell_i\in(0,\infty)$. 

\subsection{Numerical Modeling}

\begin{figure*}
	\centering
	\includegraphics[width=\linewidth]{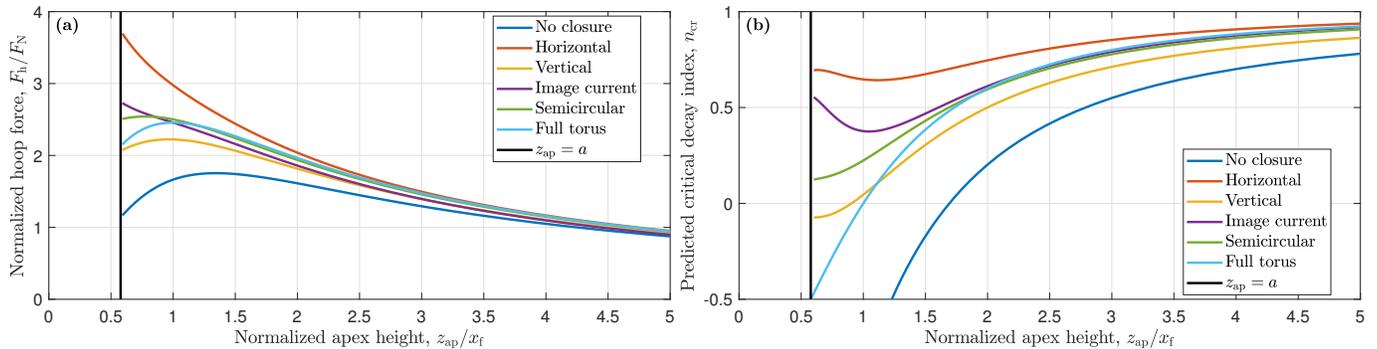}%
	\subfigure{\label{fig:decayIndex:a}}
	\subfigure{\label{fig:decayIndex:b}}
	\caption{(a) Numerically calculated hoop force and (b) critical decay index as a function of $\zapex/\xf$ for $\varepsilon=0.5$, $\ell_i=0.5$, and $n_I=0$. \Fh{} and \ncr{} have been plotted for all of the current closures shown in \Fig\ref{fig:closureModels} as well as for no closure, the unphysical situation where there are no return currents. The plots all begin at $\zapex=a$ because below this height, part of the rope would be below $z=0$ and the model breaks down. The hoop force calculations have been done numerically via the Biot-Savart Law. For plots varying the other two input parameters, $\varepsilon$ and $\ell_i$, see \Fig\ref{fig:decayIndexVsEli}.}
	\label{fig:decayIndex}
\end{figure*}

In order to study the stability of flux ropes, we investigate the force acting on a toroidal volume element around the apex, shown in gray in \Fig\ref{fig:chen-model}. Using the shifted-circle model, the hoop force acting on this region can be numerically evaluated with a simple Biot-Savart calculation. Then the normalized hoop force, $\Fh/\FN$, can be found as a function of $\zapex/\xf$, $\varepsilon$, and $\ell_i$. The hoop force has been normalized by \FNdef, and an example of this calculation is shown in \Fig\ref{fig:decayIndex:a}.

In order to find the critical decay index, \ncr, for the onset of the torus instability, first consider an MFR in equilibrium with an apex height of \zapex. The total $\bb{J}\btimes\bb{B}$ force acting on the apex can be decomposed as $F=\Fh+\Fs+\Ft$, where \Fh, \Fs, and \Ft{} are the hoop, strapping, and tension forces as defined in Table~\ref{tab:forces}. In order to find \ncr, we ask when \Eq\eqref{eqn:TI_force_balance} is marginally satisfied
\begin{equation}\label{eqn:dFdzcr}
	\left.\pD{z}{F}\right|_{F=0} \overset{!}{=} 0 \implies \pD{z}{\Fs} = -\pD{z}{}(\Fh + \Ft).
\end{equation}
The derivative of \Fs{} can be recast in terms of decay indices,
\begin{equation}
    -\frac{z}{\Fs}\pD{z}{\Fs} = -\frac{z}{\It\Bs}\pD{z}{}(\It\Bs) = n_I + \ns,
\end{equation}
where $n_I=-z\,\partial\,(\ln{\It})/\partial z$ is the decay index of the toroidal current. At marginal stability, $\ns=\ncr$, and thus the critical decay index is given by 
\begin{subequations}\label{eqn:ncr}
\begin{align} 
    \ncr &= -n_I - \left.\frac{z}{\Fh + \Ft}\pD{z}{}(\Fh + \Ft)\right\rvert_{z=\zapex}\\ 
    &= n_I - \left.\frac{z\FN}{\Fh+\Ft}\pD{z}{}\left(\frac{\Fh + \Ft}{\FN}\right)\right\rvert_{z=\zapex}, 
\end{align}
\end{subequations}
where we have also normalized all of the forces by \FN{} in the second step. This normalization of the hoop and tension forces removes their \It{}-dependency and thus isolates the effect of \It{} (and its derivatives) to the first term. In the experimental results presented in Section~\ref{sec:exp-results}, $n_I\approx0$ due to the large external inductance in the circuit driving the MFRs.

From the numerically calculated $F(\zapex)$ and \Eq\eqref{eqn:ncr}, \ncr{} can be evaluated, an example of which is shown in \Fig\ref{fig:decayIndex:b}. The numerical \ncr{} generally increases with $\zapex/\xf$ and $\varepsilon$ while it slightly decreases with $\ell_i$. The new value of \ncr{} is reduced from the standard $\ncr=1.5$ for a full torus because the general reduction of \Fh{} with height is counteracted by the emergence of a larger fraction of the torus into the $z>0$ space. Since the partial torus effects cause \Fh{} to decay slower with height than it would for a full torus, \Bs{} must also decay slower if stability is to be maintained. This reduces the value of \ncr.

\section{Experimental setup}\label{sec:exp-setup}

Flux ropes were created inside the Magnetic Reconnection Experiment (MRX) \citec{yamada1997}. The experimental setup is shown in \Fig\ref{fig:MRX} and is described in more detail in \popRef. The flux ropes were created by a discharge between two copper electrodes (ranging from $\unit[7.0-7.5]{cm}$ in radius) that were inserted into the MRX vessel with a variable footpoint half-separation of $\xf=\unit[15.5-18.0]{cm}$. The plasma was generated by injecting a small amount of neutral hydrogen ($\unit[30-40]{mTorr}$) at the vessel wall and directly at the electrodes. Breakdown was then achieved by biasing the electrodes with a capacitor bank charged to a voltage in the range of $\unit[3-4]{kV}$. In order to separate the plasma region above the electrodes from the space behind, a glass substrate is placed below the electrodes. Each discharge lasts about \unit[1]{ms}, during which time current is quasi-statically injected into the rope with a characteristic driving time of $\tau_\ms{D}\sim\unit[150]{\mu s}$, which is considerably longer than the Alfv\`en time of $\tau_\ms{A}\approx\unit[3-8]{\mu s}$. This separation of scales is what allows for MHD instabilities to be studied in MRX.

Along with the electrodes, two sets of magnetic-field coils were inserted into MRX to compliment the two external coils already present. These coils determine the configuration of the vacuum fields and are arranged such that the guide field and strapping field are independently controlled by two sets of coils each. This pairing allows for control of both the strength and decay index of the fields. The magnetic field within each rope was measured by a 2-D array of over 300 magnetic pickup coils inserted at the rope apex. By controlling the magnitude and direction of the currents in the four coils, both \qa{} and \ns{} can be selected for each shot. 

In addition to the numerical model of the hoop force discussed in Section~\ref{sec:model}, numerical corrections have been added to \Fh{} due to eddy currents that formed in the vessel wall well above the ropes at a height of $z_\ms{w}=\unit[68]{cm}$. Due to the large scale separation between the wall skin time ($\tau\sim\unit[3]{ms}$) and the flux rope driving time ($\tau_\ms{D}\sim\unit[150]{\mu s}$), the wall can be considered a perfect conductor on the timescale of the rope's lifetime. This makes calculating the mutual inductance and driven eddy currents rather straight forward. There are also eddy currents that can be driven in the center stack that runs through MRX and under the ropes. However, the skin time of the center stack ($\tau\sim\unit[75]{\mu s}$) is shorter than the driving time and therefore neglected. 

\section{Experimental results}\label{sec:exp-results}

\begin{figure}
	\centering
	\includegraphics[width=\linewidth]{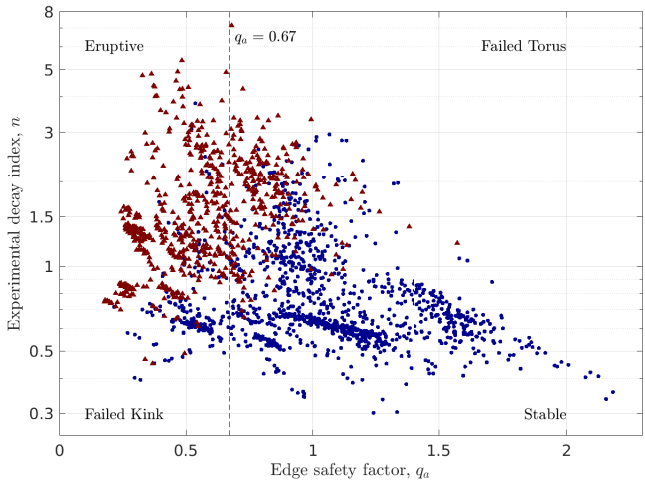}
		\caption{The experimental parameter space. Each point represents a shot and is placed based on its value of the two ideal MHD instability parameters, \qa{} and $n$. Shots that experienced multiple eruptions are represented with red triangles while non-eruptive shots are blue circles. Four different regions of stability can be seen, though their boundaries are not perfectly defined by the parameters. The dashed line at $\qa=0.67$ is an empirical cutoff used to isolate the torus instability transition from the failed torus regime. For later analysis, we will only consider the shots to the left of this line.}
	\label{fig:ns_vs_q}
\end{figure}

\begin{figure*}
    \centering
    \includegraphics[width=\linewidth]{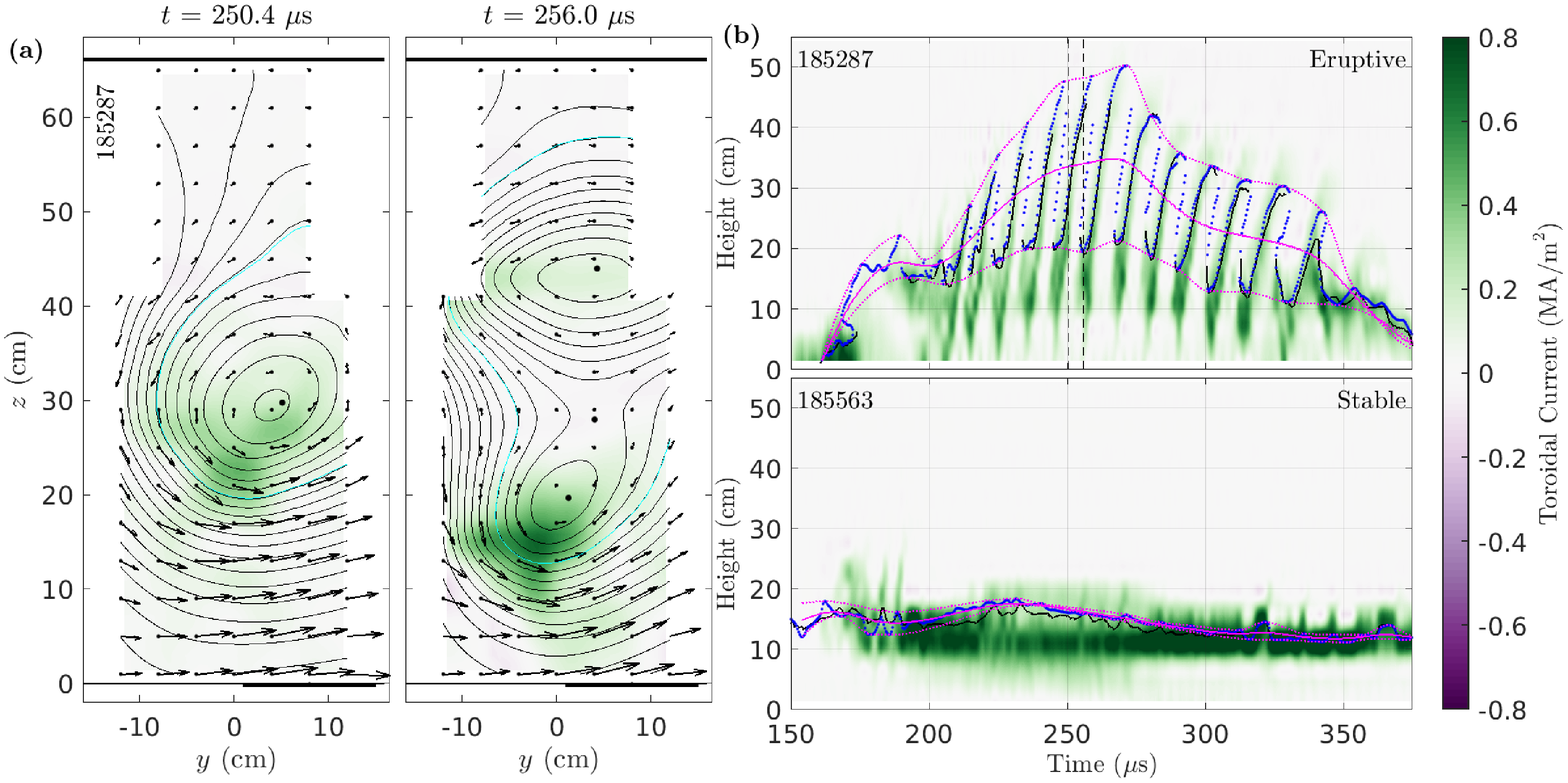}
    \subfigure{\label{fig:eruptive_J_example}}
    \subfigure{\label{fig:apex_height}}
    \caption{\textbf{(a)} Example cuts of the toroidal current density (in color) at the apex of an eruptive flux rope with arrows representing the in-plane magnetic field and dots at the nulls of this field. Contours of constant flux are shown with black curves and the flux contour that contains 80\% of the total current is marked in cyan. The first panel is a time when the rope is rising and only a single apex value exists. The second panel is a later time after a lower rope has formed that will soon erupt while the remaining portion of the initial rope can still be seen. This example event is counted as an eruption.  \textbf{(b)} An example of the time evolution of the apex height in an eruptive rope (top) compared with a stable one (bottom). The locations of the apex height are plotted with blue dots and a filtered height and envelope are plotted in magenta. The peak value of toroidal current density at each height is also plotted in color. The times shown in (a) are marked with dashed lines in the top panel. Near the end of each eruption in the eruptive example, the apex height becomes multi-valued as a new rope is formed at a lower height.}
    \label{fig:eruption_example}
\end{figure*}

In \Fig\ref{fig:ns_vs_q}, each shot taken during this campaign is plotted based on its value of the stability parameters for the kink and torus instabilities, \qa{} and $n$. The results of almost 2,000 
discharges are represented in the plot. Shots that erupted multiple times (described below) are shown by red triangles while non-eruptive shots are blue circles. The quadrants of the parameter space are labeled with their stability behavior\footnote{For a more detailed description of each quadrant see \citet{myers2016}}. We are currently interested in the onset of the torus instability and so will focus on the transition from the ``Failed Kink" to the ``Eruptive" quadrants. To this end, we will limit our analysis to those shots with a safety factor below an empirical value, $\qa<0.67$, marked in \Fig\ref{fig:ns_vs_q} as a dashed line. This value was chosen based on statistical arguments that is presented in Section~\ref{sec:statistics}. The decay index that is plotted is the decay index from the total field, \ndef, rather than just \Bs. This is done in part to be more consistent with how similar plots are made from solar observations (e.g., \citet{jing2018}).

When a rope becomes unstable and erupts, its total current would normally drop. However, there is a large external inductance in the capacitor bank driving the plasma current. This inductance prevents the total current from quickly changing, and so, instead of an erupting rope dissipating, a new rope is formed at a lower height. Since the vacuum fields have not significantly changed, the new rope is often unstable as well and quickly erupts. As an example, a cross section of the toroidal current distribution at the apex is shown for two times in \Fig\ref{fig:eruptive_J_example}. In the first panel only one rope exists, but since it is unstable, it quickly rises. Eventually a new rope is formed at a lower height, causing two distinct ropes to briefly coexist. This new rope then rises and erupts, repeating the pattern. The apex is defined by nulls in the in-plane (poloidal) field and can be multi-valued if multiple nulls exist. The poloidal field is represented by arrows and the nulls by black dots. When multiple ropes exist, there are usually 3 nulls: two O-points, one at each rope apex, and an X-point between them. An eruption was thus defined as the coexistence of two ropes, and a rope is considered ``eruptive" if this occurs more than once. An example of the repeated eruptions is shown in \Fig\ref{fig:apex_height}. This plot shows the evolution of the apex height vs. time along with a stable example for reference.

\subsection{Prediction of \ncr}\label{sec:ncr-prediction}

\begin{figure}
	\centering
	\includegraphics[width=\linewidth]{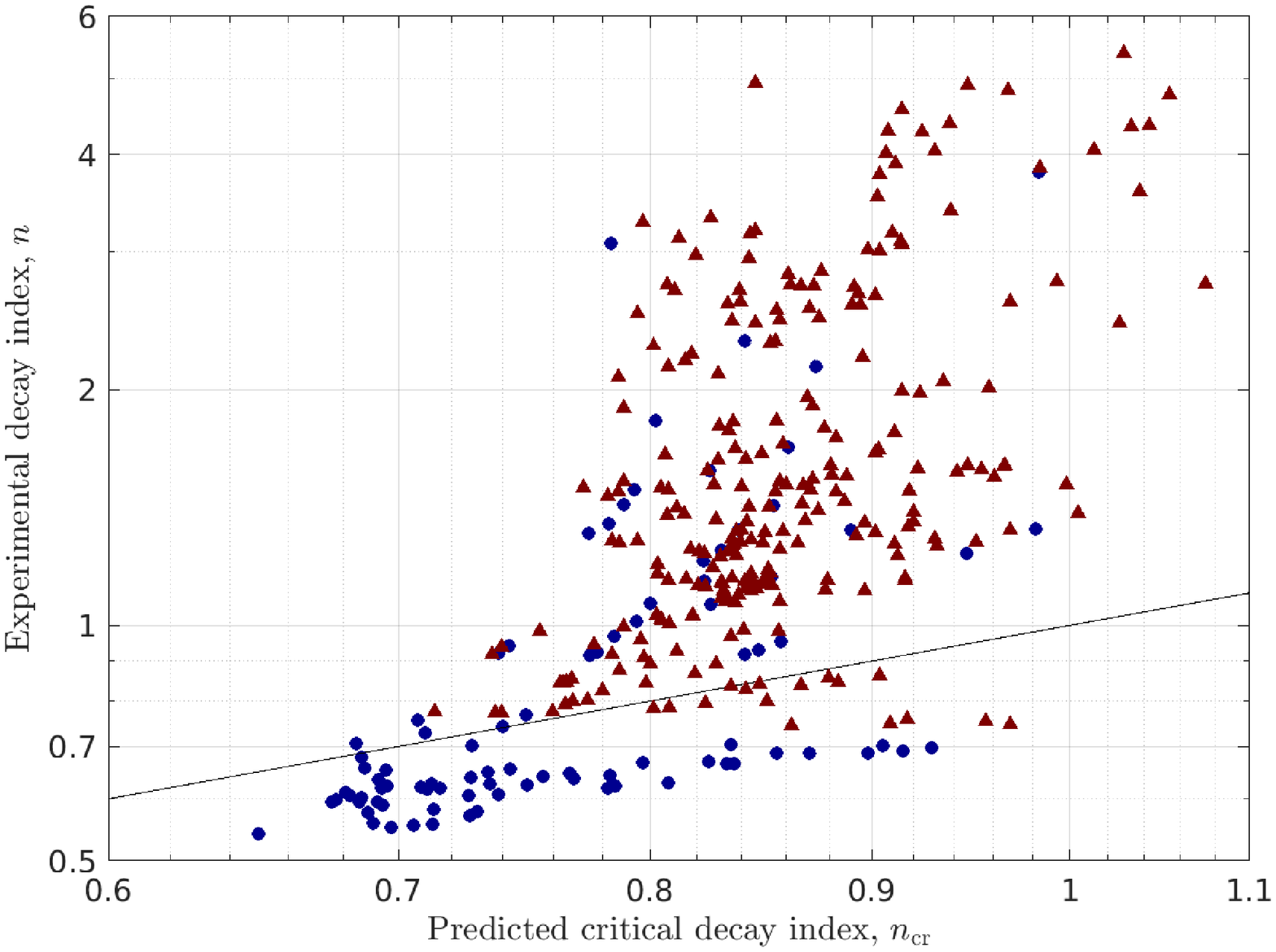}
	\caption{The measured, experimental decay index, $n$, vs. the numerically calculated, predicted critical decay index, \ncr{} presented on a log-log plot. The represented shots are limited in edge safety factor, $\qa<0.67$. Eruptive shots are represented with red triangles and non-eruptive shots with blue circles. The experimental parameters, $\zapex/\xf$, $\varepsilon\equiv a/R$, and $\ell_i$ of each shot are fit to our numerical model to determine a value of \ncr. The line $n=\ncr$ is also plotted. The model predicts that all shots above this line should be unstable while the ones below it are stable.}
	\label{fig:ns_vs_ncr}
\end{figure}

Using our more complete numerical model of the hoop force and applying \Eq\eqref{eqn:ncr}, we can predict the critical decay index, \ncr, for each shot. Each shot is fit to the model based on its experimental parameters, $\zapex/\xf$, $\varepsilon$, and $\ell_i$. The ranges of these parameters that are represented by the experimental data are discussed in Section~\ref{sec:parameter_ranges}. In \Fig\ref{fig:ns_vs_ncr} each shot is plotted based on this predicted \ncr{} and the experimentally measured $n$. In this type of plot, unstable shots should lie above the line $n=\ncr$ (also plotted) and stable ones below it. With some exceptions that will be discussed below, this is generally confirmed by the figure. This plot shows numerical evidence for the $\ncr\sim0.8$ empirical value of the critical decay index observed previously in \citet{myers2015}.

There are some caveats to \Fig\ref{fig:ns_vs_ncr} that deserve explanations. In order to get good experimental agreement, the horizontal closure from \Fig\ref{fig:closureModels} was used instead of the semicircular closure that is more representative of the experimental setup. The real experimental closure is the physical path taken by the wires leading from the footpoint electrodes to the capacitor bank and is somewhere between these two options which causes a slight overestimation of \ncr. 

There are a number of shots in \Fig\ref{fig:ns_vs_ncr} that do not exactly follow the predicted $n=\ncr$ stability line. Most of these exceptions are isolated stable shots above the line but do not alter the statistical evidence of our overall prediction. One potential cause of discrepancy comes from the value of $B$ that is used in $n$. When using the value of $B_y$ in \ns{}, it is assumed that the rope is exactly perpendicular to the measurement array. If instead, the rope's axis were  tilted at a small angle, $\theta$, relative to \ex, the nominal guide field direction, then the correct effective strapping field would be $B_\ms{s,eff} = B_y\cos\theta - B_x\sin\theta$, where $B_y$ and $B_x$ are, respectively, the strapping and guide fields when there is no tilt. Due to experimental constraints, the guide field, $\Bg$, is often considerably larger than the strapping field, $\Bs$, and so even small tilt angles can significantly change the decay index. Therefore, for robustness of our results, the total field is used here to calculate \ncr{}, which is consistent with some previous work \citec{myers2015,liu2008}. Several attempts were made to measure the tilt angle of a given rope from the available data, however nothing convincing was found. Further tests should be preformed in future experiments to better measure the tilt angle, and thus, $B_\ms{s,eff}$.

\begin{figure}
    \centering
    \includegraphics[width=\linewidth]{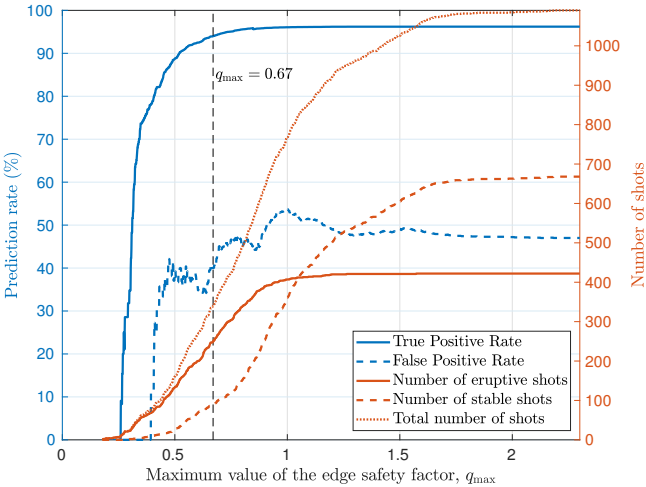}
    \caption{The statistical validity of our \ncr{} prediction as a function of the maximum value of the edge safety factor that is to be considered, \qmax. The true positive rate and the false positive rate are both plotted as functions of \qmax. The number of shots that are considered for each threshold is also plotted both as a fraction of the total and an absolute number. The value of $\qmax=0.67$ that has been used in the previous analysis has been chosen to maximize the TPR with minimal gain in FPR. It coincides with $\ms{TPR}=94\% $ and $\ms{FPR}=40\% $.}
    \label{fig:ROC}
\end{figure}

\subsection{Statistical Analysis}\label{sec:statistics}
As mentioned in Section~\ref{sec:TI-KI-effect}, we have eliminated the effect of the ``failed torus" region by only focusing on shots with an edge safety factor below a set value, \qmax. In order to investigate the effect of this choice of \qmax{}, the statistical validity of our \ncr{} prediction was compared for different values of \qmax. In \Fig\ref{fig:ROC}, we compare two metrics, the true positive rate (TPR, the fraction of eruptive shots that were correctly predicted as eruptive) and the false positive rate (FPR, the fraction of non-eruptive shots that were incorrectly predicted as eruptive). As \qmax{} is increased, the TPR quickly rises and then plateaus near 96\%. However, increasing \qmax, also causes the FPR to increase as more shots are included that are not purely eruptive and behave more like a ``failed torus". For the best results, a value of \qmax{} must be chosen to balance these two effects. By inspecting \Fig\ref{fig:ROC}, a value of $\qmax=0.67$ was chosen because for larger values, the FPR continues to rise for little gain in the TPR. This value of \qmax{} results in values of $\ms{TPR}=94\% $ and $\ms{FPR}=40\% $. It should also be noted that the total number of shots represented in \Fig\ref{fig:ROC} is about half that of \Fig\ref{fig:ns_vs_q}. This is because a number of shots were taken with the measurement array not perpendicular to the MFR axis. The angle removes our ability to predict \ncr{} and thus they cannot be included in the statistics.

\subsection{Experimental Parameter Ranges}
\label{sec:parameter_ranges}
\begin{figure}
    \centering
    \includegraphics[width=\linewidth]{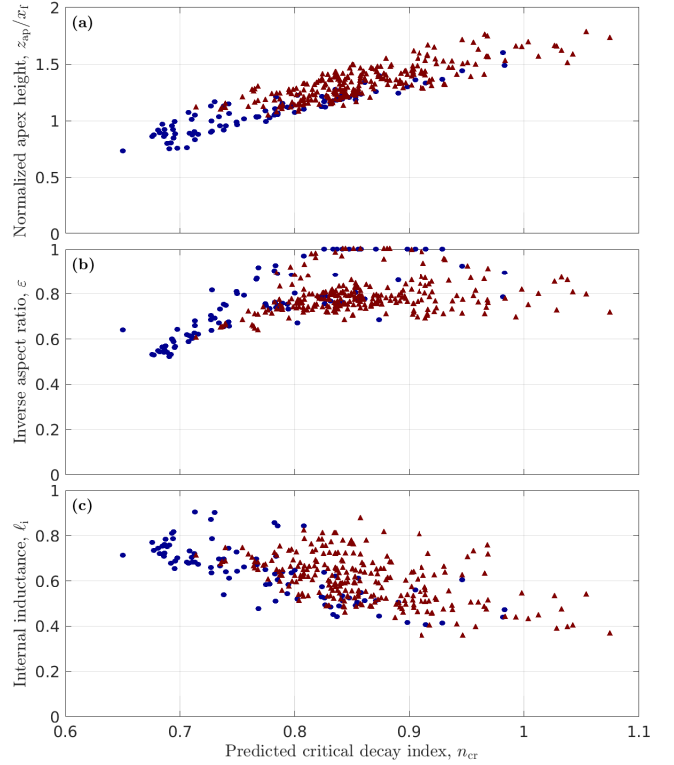}
    \subfigure{\label{fig:params-vs-ncr:a}}
	\subfigure{\label{fig:params-vs-ncr:b}}
	\subfigure{\label{fig:params-vs-ncr:c}}
    \caption{A scatter plot of the dimensionless, experimental parameters used in the calculation of \ncr{}, (a) $\zapex/\xf$, (b) $\varepsilon$, and (c) $\ell_i$ compared to the predicted \ncr{} value. The shots represented here are the same as in \Fig\ref{fig:ns_vs_ncr} and are limited in edge safety factor, $\qa<0.67$.}
    \label{fig:params-vs-ncr}
\end{figure}

As discussed in Section~\ref{sec:ncr-parameters}, the calculation of a predicted \ncr{} involves three dimensionless experimental parameters: the normalized apex height, $\zapex/\xf$, the inverse aspect ratio, $\varepsilon$, and the internal inductance, $\ell_i$. The shots shown in \Fig\ref{fig:ns_vs_ncr} define ranges of these parameters based on experimental constraints. Scatter plots of these values vs. the predicted \ncr{} for each of these shots are shown in \Fig\ref{fig:params-vs-ncr}. The parameters fall within the ranges of $\zapex/\xf\in(0.75,1.75)$, $\varepsilon\in(0.5,1.0)$, and $\ell_i\in(0.3,0.9)$. Care should be taken when extrapolating our results beyond these values. Some correlations can also been seen between the parameters in these plots and \ncr. The strongest of these correlations is between $\zapex/\xf$ and \ncr{}, which is not surprising based on the predicted \ncr{} curves presented in \Fig\ref{fig:decayIndexVsEli}.

It should be noted that some shots had a measured $\varepsilon>1$. This is possible due to varying minor radius over the length of the rope, which is not captured by our model. Since, especially at large $\varepsilon$, there is only minor changes in \ncr{} with $\varepsilon$ (see \Fig\ref{fig:decayIndexVsEli:a}), these shots have had their aspect ratio adjusted to $\varepsilon=1$.

\section{Application to solar flux ropes}

\begin{figure*}
	\centering
	\includegraphics[width=\linewidth]{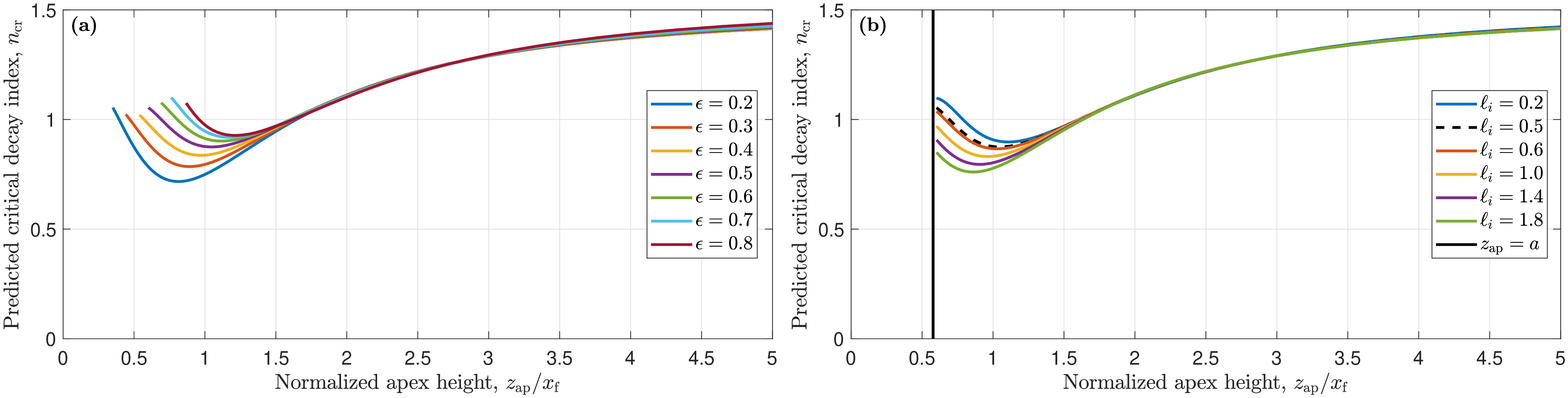}
	\subfigure{\label{fig:decayIndexVsEli:a}}
	\subfigure{\label{fig:decayIndexVsEli:b}}
	\caption{Numerically predicted critical decay indices for solar applications as a function of normalized apex height, plotted for different values of $\varepsilon$ and $\ell_i$. The plots here are chosen to be most applicable to solar conditions with an image current closure and toroidal current decay index of $n_I=0.5$. (a) $\ncr(\zapex/\xf)$ for different values of $\varepsilon$ at fixed $\ell_i=0.5$. (b) $\ncr(\zapex/\xf)$ for different values of $\ell_i$ at fixed $\varepsilon=0.5$. The minimum value of $\zapex/\xf$ where the MFR would pass below $z=0$ is also shown for this plot.}
	\label{fig:decayIndexVsEli}
\end{figure*}

\begin{deluxetable*}{ lp{40mm}p{40mm}p{50mm} }
	\tablecaption{Qualitative comparison of experimental and solar conditions with resulting \ncr{} adjustment\label{tab:solarComparison}}
	\tablehead{
	\colhead{Parameter}     &   \colhead{Solar value}   &   \colhead{Laboratory value}  &  \colhead{\ncr{} adjustment}} %
	\startdata
    \raggedright Current return path   &  Image currents on the solar surface   &   Fixed wires  &  Decrease of \ncr{} for low-lying ropes; $\ncrSol\approx\nolinebreak\ncrLab-0.3$%
    \\
    \hline
    \raggedright Lower boundary condition    &   Conductive surface at $z=0$; Line-tied everywhere    &   Conductive, line-tied footpoints; dielectric elsewhere%
     & Affects the poloidal flux conservation, but no direct effect on \ncr %
    \\
    \hline
    \raggedright Total toroidal current &  Set by poloidal flux conservation; $n_I\sim\nolinebreak0.5$  &  External inductance forces constant current; $n_I\approx 0$   &  Increase of \ncr{} due to $n_I$%
    ; $\ncrSol=\nolinebreak\ncrLab+n_I\sim\nolinebreak\ncrLab+0.5$%
    \\
    \hline
    \raggedright Upper boundary condition   &   Open boundary at $z\to\infty$ allows ropes to erupt away  &   Conductive wall prevents ropes from fully erupting  &  Minor effects except in post-eruption stages; $\ncrSol\approx\nolinebreak\ncrLab-0.1$ \\
    \enddata
\end{deluxetable*}

The laboratory flux ropes we have created are most directly comparable to solar flux ropes with some considerations that are laid out in Table~\ref{tab:solarComparison}. There are three main differences that we will focus on: (1) the evolution of the total toroidal current; (2) the sub-surface closure of the current; and (3) eddy currents in the outer vessel wall. 

The total toroidal current as a rope rises behaves very differently on the Sun when compared to the laboratory case. In the laboratory, the total plasma current is fixed due to the large external inductance in the driving circuit. However, on the Sun the current is determined by the conservation of magnetic flux between the rope and the solar surface. This affects \ncr{} through the current decay index, $\ncrSol=\ncrLab+n_I^\text{solar}$. The value of $n_I$ depends on the model being used, but for the standard large-aspect-ratio, full torus, $n_I^\text{solar}\approx0.5$. This value is also similar to that used in other models. When the toroidal current is not able to decrease with height, the MFRs are less stable to the torus instability. %

The laboratory return path of the current through fixed wires rather than through a conductive surface as on the Sun also has an effect on \ncr. The effect is largest on low-lying ropes, and when the changes between the models are compared, we find that \ncr{} is decreased by about $-0.3$. %

\begin{figure}
	\centering
	\includegraphics[width=\linewidth]{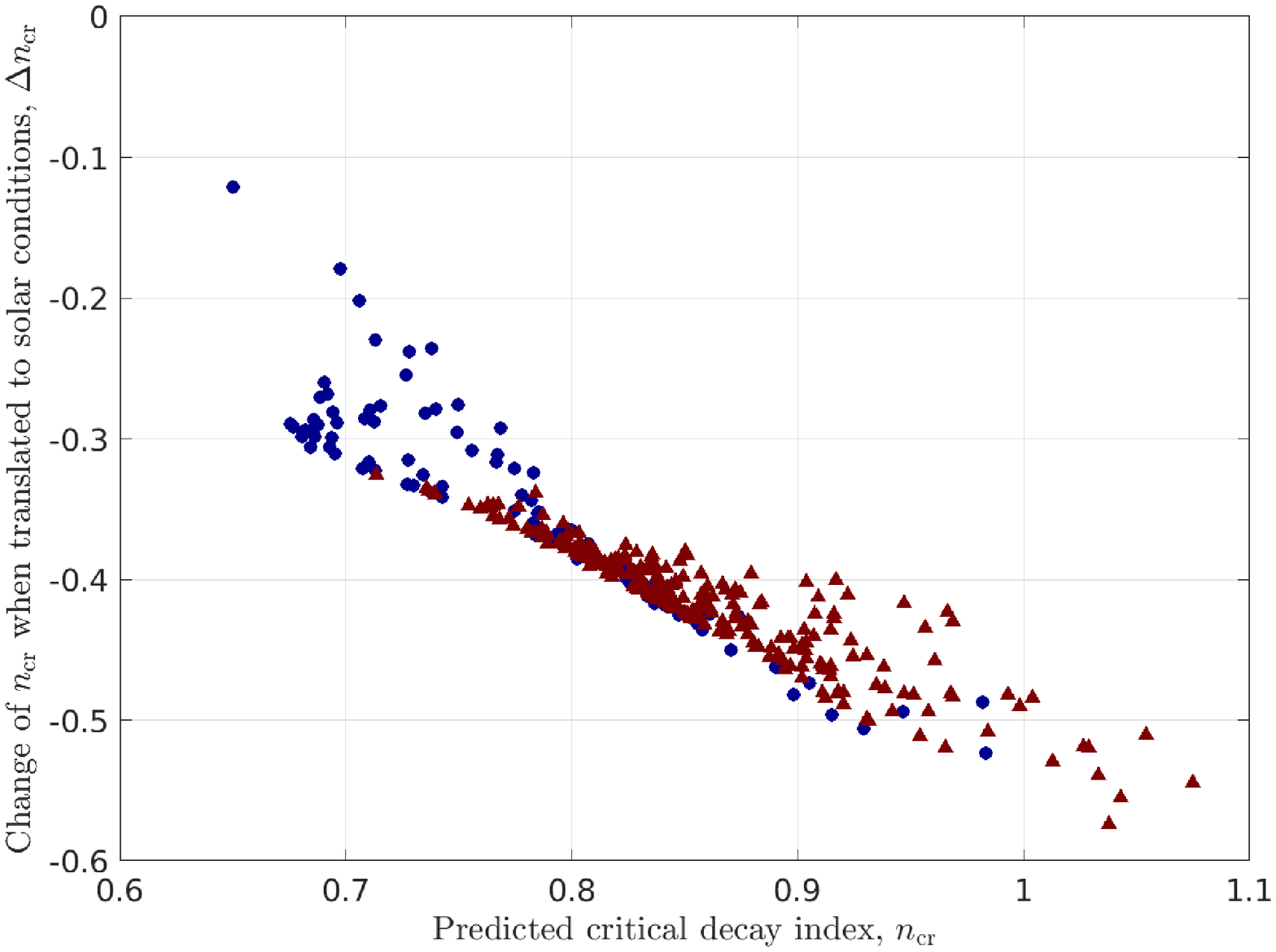}
	\caption{The change in the predicted \ncr{} when the laboratory conditions are translated to the Sun but with $n_I=0$. Each shot in \Fig\ref{fig:ns_vs_ncr} is represented with eruptive shots again being represented by red triangles and non-eruptive shots by blue circles. The changes due to current closure and eddy currents in the outer vessel wall are both accounted for and create a change centered on $\Delta\ncr\sim-0.4$ in a larger range of $\Delta\ncr\sim(-0.5,-0.3)$.}
	\label{fig:dncr_vs_ncr}
\end{figure}

The effect of eddy currents in the outer wall located at $z\approx3.8\xf$ has also been numerically calculated. The wall has the largest effect on tall ropes, decreasing \ncr{} by about $-0.1$. The effect of the wall is much larger on ropes that would fully erupt away. However, since we are focused on the onset of eruption and not the dynamics afterward, this only has a minimal effect on our results. The specific change in \ncr{} due to these last two effects is shown for each shot in \Fig\ref{fig:dncr_vs_ncr}.

The net effect on \ncr{} when translating from laboratory to solar conditions is therefore 
\begin{equation}
    \ncrSol\approx\ncrLab+n_I-0.4\sim\ncrLab+0.1 .
\end{equation}
The modified \ncrSol{} is shown in \Fig\ref{fig:decayIndexVsEli}. These plots show \ncrSol{} as a function of $\zapex/\xf$ for various values of $\varepsilon$ with fixed $\ell_i=0.5$ and then for various values of $\ell_i$ with fixed $\varepsilon=0.5$. These 1-D cuts help in breaking down the three dimensional input space. It can be easily seen that the effects of changing either $\varepsilon$ or $\ell_i$ is very minimal for tall ropes with $\zapex/\xf>1.5$ but can be important for low lying ropes. From these cuts, it is seen that \ncrSol{} increases with increasing $\varepsilon$ and decreases with increasing $\ell_i$. That is, MFRs with lower aspect ratios and less peaked currents are more stable to the torus instability. The non-monotonic curves in \Fig\ref{fig:decayIndexVsEli} are caused by two competing effects as the MFR rises: \Fh{} decreases overall and a larger fraction of the torus is present, increasing \Fh. The latter effect becomes smaller as the MFR rises since the relative increase is less.

Another difference between the laboratory and solar conditions is the state of the volume surrounding a flux rope. On the Sun, the external volume is the ambient plasma of the corona with frozen-in flux. However, in our experiments, the plasma is generated by the arc discharge that forms the rope and so most of the external volume is near vacuum with small neutral gas pressure. On the Sun, an erupting rope must displace this ambient plasma away while there is no such force in our experiments. It has been seen in simulations that when this background plasma contains a large guide field, otherwise unstable ropes can be stabilized \citec{kliem2014}. However, we have not seen this effect in our experiments, possibly because of the lack of ambient plasma. Since we are currently focusing on ropes with small \qa{}, the ambient guide field is also small, and this effect likely would not be applicable.

\section{Discussion and conclusions}

The onset criteria of the torus instability in arched, line-tied flux ropes has been investigated by laboratory experiments and the application of a simplified theoretical model. Arched, line-tied flux ropes are modeled by a shifted circle with fixed footpoints (see \Fig\ref{fig:closureModels}) and the resulting forces are derived from simple Biot-Savart calculations. The effect on the torus instability of a partial torus model has been previously considered in \citet{olmedo2010}. However, the use of numerical calculations in the hoop force of a partial torus allows for more accurate predictions of its onset and for the application to low aspect ratio tori with small edge safety factors. 

When applied to our experimental data, our model has predicted critical decay index values within the range of $\ncrLab\sim(0.65,1.1)$ which is consistent with the previous empirical value of $\ncrLab\sim0.8$ \citec{myers2015}. Thus our new model provides quantitative evidence for the previous empirical description of the flux rope stability regimes. The deviation of this range from the standard value of 1.5 \citec{kliem2006}, is caused by taking into account the full effects of the partial torus when applied to the conditions of the laboratory experiments. A large portion of the decreased \ncrLab{} is due to the external inductance in the flux rope driving circuit which prevents the total plasma current from changing. While this is a major change from solar conditions, its effect is easily quantified and separated, allowing for direct comparisons. Further changes are caused by the modification of the hoop force from partial torus effects. In particular, the emergence of more of the torus as the rope rises counteracts some of the decrease of \Fh{}. %

While the predicted \ncrLab{} largely agrees with our experimental results there are still a number of exceptions that were discussed in Section~\ref{sec:ncr-prediction}. These exceptions can possibly be explained by a small titling of the flux ropes relative to the measurement array. However, since this angle was not able to be found through the measurements taken, there may of course be other explanations that have yet to be considered. Further investigation into this portion of the parameter space may yield a better understanding of these ropes and their unexpected behavior. Even with these exceptions, our model is able to predict eruption ropes with a true positive rate of 94\%. This rate of prediction is very substantial when considering the complexity of MFR eruptions. There is still a relatively high false positive rate of 40\%. However, for application to space weather prediction, false positives are better than false negatives as they would represent extra caution rather than missed events that could cause irreparable damage.

Our analysis has also assumed small values of \qa{} to eliminate the ``failed torus" regime. The twist number is difficult to measure in solar observations, but the maximum value we have chosen, $\qmax=0.67$ (or equivalently $|T_\ms{w,min}|=1/\qmax=1.49$), is relevant to many solar MFRs \citec{duan2019}. When translated to solar conditions, we find that most of our predicted critical decay index values fall near $\ncrSol\sim0.9$ and within a larger range of $\ncrSol\sim(0.7,1.2)$, which is consistent with some solar observations \citec{jing2018}. This range provides a more accurate description of \ncrSol{} and our methodology can help explain the discrepancies seen in previous numerical and observational studies. This new methodology allows for the novel application of laboratory experiments directly to the conditions present on the Sun. The ability to perform direct, local measurements in the laboratory gives experiments a unique advantage over other methods of studying solar flux ropes.

\section*{Acknowledgements}

We would like to thank P. Sloboda, A. Jones, J. Latham, L. Bevier, and R. Cutler for their help with the experimental setup. This research is supported by Department of Energy contract numbers DE-SC0019049 and DE-AC0209CH11466 (Princeton) and NASA grant numbers 80HQTR17T0005, 80NSSC17K0016, and 80NSSC19K0860. Sandia National Laboratories is a multimission laboratory managed and operated by National Technology \& Engineering Solutions of Sandia, LLC, a wholly owned subsidiary of Honeywell International Inc., for the U.S. Department of Energy’s National Nuclear Security Administration under contract DE-NA0003525. This paper describes objective technical results and analysis. Any subjective views or opinions that might be expressed in the paper do not necessarily represent the views of the U.S. Department of Energy or the United States Government.

\bibliography{MFR-bibliography}

\end{document}